\documentclass[prb,superscriptaddress,twocolumn]{revtex4-2}
\usepackage{graphicx}
\usepackage{times,amsmath}
\usepackage{epsfig}
\usepackage{color}
\usepackage{graphicx}
\usepackage{dcolumn}
\usepackage{bm}
\usepackage{bookmark}
\usepackage{tabularx}
\usepackage{hyperref}
\usepackage{multirow}
\usepackage{array,mathtools,amssymb,booktabs,makecell}
\hypersetup{colorlinks=true, citecolor=blue, filecolor=blue, linkcolor=blue, urlcolor=blue}

\usepackage{float}
\usepackage{adjustbox}
\usepackage{upgreek}
\makeatletter
\let\saved@includegraphics\includegraphics
\AtBeginDocument{\let\includegraphics\saved@includegraphics}
\renewenvironment*{figure}{\@float{figure}}{\end@float}
\makeatother

\begin{document}


\title{Mott transition and abnormal instability of electronic structure in FeSe}

\author{Byungkyun Kang}
\email[]{bkang@udel.edu}
\affiliation{College of Arts and Sciences, University of Delaware, Newark, Delaware 19716, USA}

\author{Maengsuk Kim}
\affiliation{Quantum Matter Core-Facility and Research Center of Dielectric and Advanced Matter Physics, Pusan National University, Busan 46240, Republic of Korea}

\author{Chul Hong Park}
\affiliation{Quantum Matter Core-Facility and Research Center of Dielectric and Advanced Matter Physics, Pusan National University, Busan 46240, Republic of Korea}

\author{Anderson Janotti}
\affiliation{Department of Materials Science and Engineering, University of Delaware, Newark, Delaware 19716, USA}

\begin{abstract}
FeSe has been extensively explored as a quantum material, primarily due to the observed highest superconducting transition temperature among Fe-based 
unconventional superconductors. Nonetheless, the electronic structure and the electron correlations responsible for the remarkable diversity of physical properties in FeSe remain elusive.
We undertook a comprehensive investigation of the electronic structure of FeSe, known as a Hund metal,
and found that it is not uniquely defined.
Through accounting for all two-particle irreducible diagrams constructed from electron Green's function $G$ and screened Coulomb interaction $W$ in a self-consistent manner, a Mott-insulator phase of FeSe is unveiled.
The metal-insulator transition is driven by the strong on-site Coulomb interaction in its paramagnetic phase, accompanied by the weakening of both local and non-local screening effects on the Fe-3$d$ orbitals.
Our results suggest that Mott physics may play a pivotal role in shaping the electronic, optical, and superconducting properties of monolayer or nanostructured FeSe.


\end{abstract}

\maketitle

\textit{Introduction.} In recent years, quantum materials have taken the forefront in materials science, revealing unique electronic and magnetic properties that arise from intricate quantum mechanical effects. This field has the potential to revolutionize a wide range of applications, including electronics, quantum computing, and energy technologies\cite{keimer_natphys2017}.
Iron-based superconductors (IBSs) 
have attracted much attention due to the potential connection to high-temperature cuprate superconductors\cite{yoichi_jacs2008,ren_cpl2008,marianne_prl2008,fong_pnas2008,wang_ssc2008,defa_ncom2012}.
Among them, FeSe has emerged as a fascinating material due to its unusual phase diagram, which includes bad metallicity arising from Hund's correlations\cite{antonie_arcmp2013,yin_prb2012}, a superconducting state near the boundary of orbital-selective Mott phase\cite{yi_prl2013}, as well as its potential for applications in quantum devices and topological quantum computing\cite{camron_prl2023}.
Captivatingly, the high-T$_c$ superconductivity up to 80 K in monolayer FeSe on perovskite oxides has been measured\cite{jian_nmat2014,shaolong_nmat2013}.
To grasp the mechanism of this exceptional superconductivity, a comprehensive understanding of the electronic structure of FeSe becomes imperative.

As evidenced by the superconductivity and the ARPES measurement and the theoretical calculations\cite{yi_ncom2015,laura_prb2016,yin_prb2012,matthew_prb2017}, FeSe has been identified as a metal in contrast to the high-temperature superconducting cuprates whose parent compounds are Mott insulators.
However, some recent work of the thin-film or monolayer FeSe on perovskite oxides such as SrTiO$_3$ and BaTiO$_3$ indicated the possibility that insulating and metallic states coexist\cite{wang_nature2016,hanzawa2019insulator}. 
A fascinating report is that the superconducting temperature was shown to be controllable using MOSFET structure by electric field. When the gate voltage was low, the FeSe showed semiconducting behavior with low carrier concentration\cite{hanzawa2016electric}.
In addition, a recent measurement of photo-luminescence of FeSe nanoparticles\cite{park_nono2022} has raised questions about the electronic structure of FeSe.
FeSe is classified as a strongly correlated 2-dimensional quantum material\cite{amalia_arcmp2018}, and a thorough understanding of strong electron correlations is necessary for a comprehensive understanding of its inconsistent behavior.

Since strong electron-electron interaction can prevent Cooper-pair formation, an unconventional superconductivity mechanism should work for IBSs\cite{rong_ncom2013}.
The superconductivity generally appears in the vicinity of the antiferromagnetic phase. However, FeSe holds a unique position among IBSs, as bulk FeSe has no magnetic ordered phase, unlike most other materials\cite{fong_pnas2008}. 
A recent study\cite{chang_npjcm2020} suggests that suppressed magnetic order in bulk FeSe is associated with the reduction of interorbital charge ﬂuctuations, induced by Hund's coupling and large crystal-field splitting. In contrast, the reduced dimensionality and spatial isolation of Fe atoms in the expanded monolayer FeSe result in a strong magnetic order. 

The electronic structure of Fe-3$d$ states 
is strongly affected by the ligand field from the local crystal structure. Since Fe is in 2+ oxidation state in FeSe, only one of the Fe-3$d$ levels of the minority spin is occupied. The Fe-3$d$ levels are degenerate, either in the t$_{2g}$ or e$_g$ levels, depending on the ligand field and the local symmetry.
The electronic structure should be sensitive to the angle around Fe. Since the Fe is 4-fold-coordinated, the local symmetry is believed to be close to tetrahedral. However, the angle around Fe is quite different from the ideal tetrahedral angle. In the 2-dimensional structure, the lattice relaxation energy is small, by which the effect of Jahn-Teller (JT) lattice distortion is expected to affect the electronic structure significantly.
Lattice distortion has been shown to influence electron-electron correlations, which in turn can affect superconductivity.

\begin{figure}[ht]
\centering
\includegraphics[width=0.50 \textwidth]{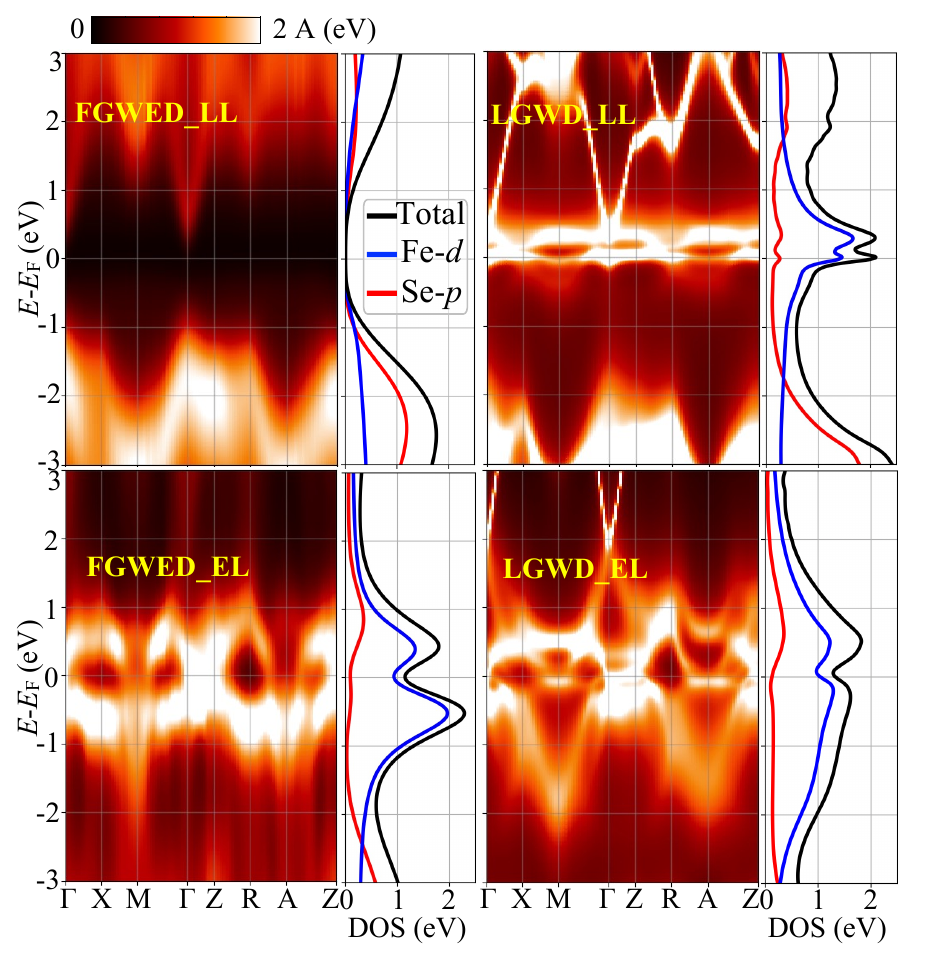}
\caption{\label{Fig_band}Calculated spectral functions, the total density of state (DOS), and the atomic orbital projected DOS with the full GW+EDMFT (FGWED) and the LQSGW+DMFT (LGWD) for large (LL) and experimental (EL) lattices are shown. 
}
\end{figure}

It has been observed that the T$_{c}$ of unconventional superconductor FeSe is highly dependent on the lattice structure.
Co-doping of S/Te on the Se-site has been shown to have no effect on the lattice parameter. However, it has been observed to increase the density of electronic states (DOS) near the Fermi level, thus enhancing the superconductivity\cite{fan_jac2017}.
Nonetheless, the doping of S or Te on Se-site in FeSe resulted in changing lattice and increasing T$_{c}$~\cite{mizu_jpsc2009}. In FeSe$_{1-x}$Te$_{x}$, both lattice parameters $a$ and $c$ increased in accordance with the ionic radius of Te. Conversely, in FeSe$_{1-x}$S$_{x}$, both $a$ and $c$ decreased due to smaller ionic radius of S compared to that of Se.
In addition, the intercalation of alkali and alkaline-earth metals into FeSe layers has been observed to cause an increase in the T$_{c}$ as well due to modification of the interlayer distance\cite{guo_prb2010,ying_scirep2012,ying_jacs2013,liu_advsci2016,guo_ncom2014,swagata_npjqm2023}.
It has been suggested that interface-strain induced spin density waves and doping effects are essential for the superconductivity observed in FeSe/SrTiO$_{3}$ thin films\cite{shiyong_namt2013}. 
However, Mandal {\em et al.}\cite{subhasish_prl2017} proposed that electronic correlations have a significant influence on the electronic structure of the monolayer FeSe on SrTiO$_{3}$. The Se-Fe-Se angle in both freestanding FeSe monolayer and FeSe/SrTiO$_{3}$ has been found to govern the electronic correlations, thus emphasizing the significance of FeSe atomic geometry. 
These observations suggest that the electronic structure of FeSe is flexible, which may be due to structural distortion\cite{yeh_epl2008,mizu_apl2009} or differences in electronegativity of chalcogens\cite{fan_jac2017} upon doping.

\begin{figure}[ht]
\centering
\includegraphics[width=0.50 \textwidth]{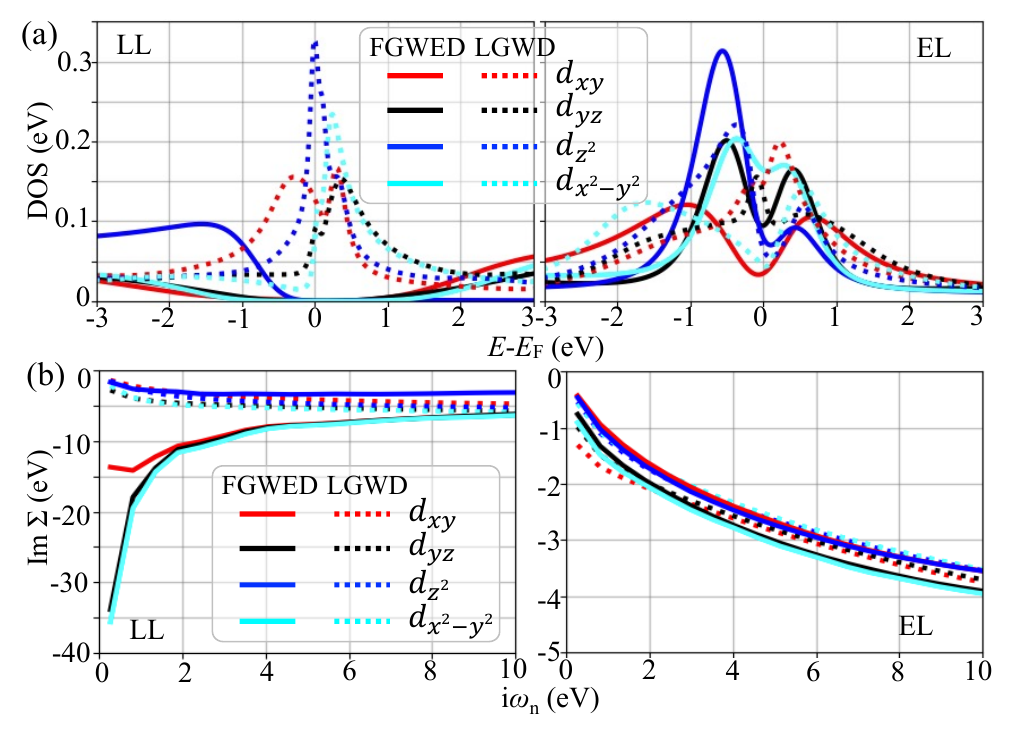}
\caption{\label{Fig_self}DOS and imaginary part of self-energies for Fe-3$d$ orbitals within full GW+EDMFT and LQSGW+DMFT for large and experimental lattices.}
\end{figure}

Due to the strong correlations, the electronic structure of FeSe is suggested to be largely determined by local on-site Coulomb repulsion $U$ and Hund's coupling $J$. This is evidenced by the results of several studies utilizing dynamical mean field theory (DMFT) combined with density functional theory (DFT)\cite{amalia_arcmp2018,antonie_arcmp2013,chang_npjcm2020,haul_njp2009,skornyakov_prb2017,markus_prb2010,matthew_prb2017,subhasish_prl2017,swagata_npjqm2023,swagata_sym2021,yin_natphy2014,yin_prb2012}.
The DFT+DMFT and angle-resolved photoemission spectroscopy (ARPES) revealed the presence of Hubbard bands of Fe-3$d$ at high binding energies\cite{matthew_prb2017,markus_prb2010}.
The DFT+DMFT calculations proposed renormalized quasi-particle bands in the vicinity of the Fermi level, which are governed by the physics of Hund metals\cite{swagata_sym2021,subhasish_prl2017,byung_lanio2,haul_njp2009,chang_npjcm2020,antonie_arcmp2013,yin_prb2012}.
Skornyakov {\em et al.} demonstrate a marked orbital-dependent renormalization of Fe-3$d$ electrons when the lattice is expanded, as determined by the DFT+DMFT approach\cite{skornyakov_prb2017}. 
This suggests that the local correlation should be linked to the aforementioned flexible lattice of FeSe upon doping or lowering dimension. However, within the conventional DFT+DMFT methods, $U$ is a parameter that is independent of the specific system, making it difficult to accurately determine electron correlations.

\begin{table}[]
\caption{Calculated  electron occupation of Fe-3$d$ orbitals. The quasi-particle weight $Z$ factors were presented in parentheses.  }\label{table_occ}
\scriptsize
\begin{center}
\scalebox{1}{
\begin{ruledtabular}
\vspace*{5mm}
\begin{tabular}{ccccc}
 \multicolumn{1}{c}{} &
 \multicolumn{2}{c}{full GW+EDMFT} &
 \multicolumn{2}{c}{LQSGW+DMFT} \\
 \hline
  $ $ & \multicolumn{1}{c}{Large LC} & \multicolumn{1}{c}{Exp. LC} & \multicolumn{1}{c}{Large LC} & \multicolumn{1}{c}{Exp. LC} \\
  \hline
  $d_{xy}$ &   1.02   &    1.22 (0.52)    &    1.40 (0.40)    &    1.22 (0.58)    \\
  \hline
  $d_{yz}$& 1.02 & 1.12 (0.49) & 1.12 (0.32) & 1.18 (0.52)\\
  \hline
  $d_{z^2}$ & 1.94 & 1.58 (0.49) & 1.26 (0.33) & 1.40 (0.49) \\ 
  \hline
    $d_{xz}$ &1.02& 1.12 (0.49) & 1.12 (0.32) & 1.18 (0.52)  \\ 
  \hline
    $d_{x^2-y^2}$ &1.02& 1.24 (0.49) & 1.08 (0.25) & 1.28 (0.52)  \\   
  
\end{tabular}
\end{ruledtabular}}
\end{center}
\end{table}

\begin{figure*}[ht]
\centering
\includegraphics[width=0.99
\textwidth]{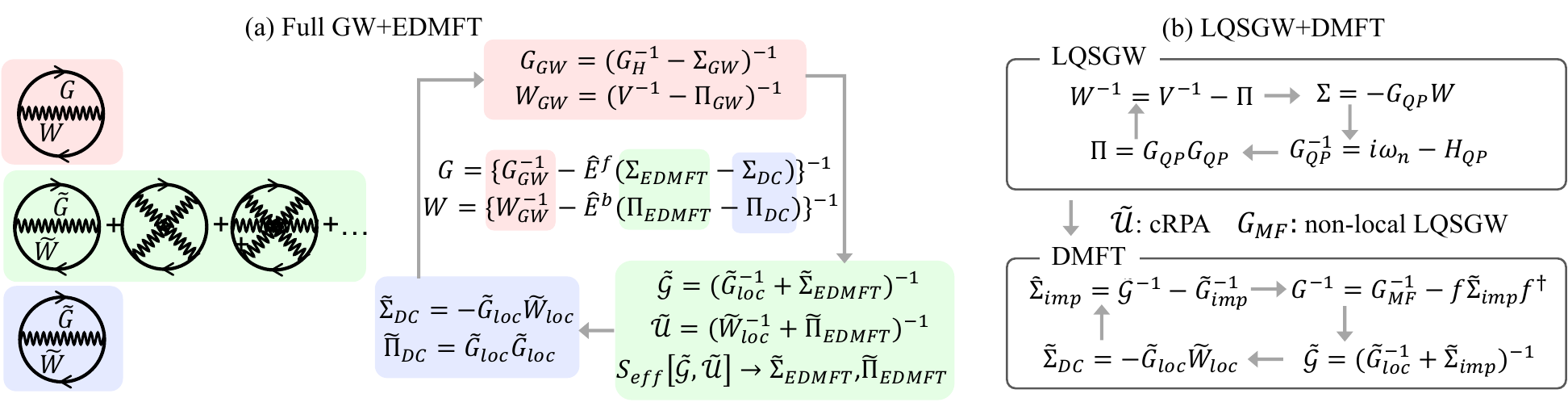}
\caption{\label{Fig_diagram} (a) Diagrammatic representations of the $\Psi$ functional (left) and theoretical framework with self-consistent loop (right) for full GW+EDMFT approach. The corresponding representations and processing part of GW $\Psi$  functional (red), EDMFT $\Psi$ functional (green), and local-GW $\Psi$ functional (blue) are marked with the same color. (b) LQSGW+DMFT framework. First, a mean-field Green’s function ($G_{MF}$) and  bosonic Weiss field $\widetilde{\mathcal{U}}$ from constrained random phase approximation (cRPA) are calculated within non-local LQSGW. Then, using $G_{MF}$ and $\widetilde{\mathcal{U}}$, DMFT equation is solved with self-consistent loop.
}
\end{figure*}

\textit{Instability of electronic structure within DFT+U.}
We first investigated the electronic and lattice structure of FeSe with long-range magnetic order, using static $U$ within DFT+$U$, which can describe the magnetic ground-state of Mott insulators.
The measured equilibrium crystal structure of the tetragonal phase (P4/nmm) FeSe\cite{joshua_cm2014} is depicted in supplementary Fig. 1.
Here, Fe is 4-fold coordinated, thus it is expected to be a tetrahedral structure with four Se atoms bound to Fe.

The electronic configuration of Fe is 3$d^6$ and 4$s^2$, which gives two electrons to Se. As described in supplementary Fig. 1, the lower levels of minority spin should be degenerate either for tetrahedral or octahedral structure.
Therefore, it is expected that the degeneracy can be removed by a JT distortion. The lattice distortion energy is usually expected to be smaller in two-dimension lattice than in a 3D lattice. Thus the JT distortion can be significant, which can also change the electronic structure.



We investigated the possibility of various electronic structures. 
Remarkably, three kinds of electronic structures are found to be available from the same type of atomic structure through LSDA+$U$ calculations.
The electronic structures for the antiferromagnetic spin-antiparallel state is described by the DOS in supplementary Fig. 2. The lattice distortions are described in supplementary Table II.
We found that the band gap can be opened by the lattice distortion and the Hubbard-$U$ interaction between occupied and unoccupied states. 
The electronic structures can be changed by breaking of spin symmetry making long-range order in the LSDA+$U$ framework.
We furthermore explore the feasibility of computing electron correlations and their quantum effects in the paramagnetic phase of FeSe by using two recently developed state-of-the-art ab initio many-body physics approaches:
i) the fully self-consistent GW+EDMFT (FGWED)\cite{kangfgwedmft} and ii) the linearized quasi-particle self-consistent GW (LQSGW)+DMFT (LGWD)\cite{choi2019comdmft}.
The LQSGW+DMFT has been utilized to capture strong electron correlations in various systems, including transition metal oxide, actinide-and lanthanide -based compounds\cite{siddiquee2022breakdown,kang2023dual,kang2022orbital,kang2022tunable,kang2023infinite,kang2019nio}.

Recent study indicated that nanoparticles of FeSe can exhibit Mott-insulating behavior and emit an unusual luminescence from the gap\cite{park_nono2022}.
The insulating state emerged by expanding the lattice parameters.
Here, we investigate the Mott transition of FeSe by comparing the experimental lattice parameters (EL) of $a=$ 3.77 and $c=$ 5.52 $\textrm{\AA}$\cite{joshua_cm2014} to large lattice parameters (LL) of $a=$ 4.25 and $c=$ 6.99 $\textrm{\AA}$. The inter-layer distance is extended by $\sim$0.7$\AA$ along the $c$-axis from an optimized lattice of spin polarized LSDA+$U$ method, using $U$ = 5 eV. Here,
the electronic structure is little changed with respect to the variation of inter-FeSe-layer distance, indicating that the FeSe layers are well separated\cite{park_nono2022}. 
Thus, this lattice can facilitate the understanding of the physics of monolayer FeSe.
It is remarkable that the latter (large lattice) insulating state has a lower total energy than the former metallic state (experimental lattice) within LSDA+$U$ method.
It was observed that the tensile stressed monolayer FeSe on SrTiO$_3$ (ML-FeSe/STO) exhibits high-temperature superconductivity\cite{jian_nmat2014,shaolong_nmat2013}, where the $a$-axis lattice constant of SrTiO$_3$ is 3.905 $\textrm{\AA}$\cite{schm_ac2012}.


\textit{Mott transition.}
A salient finding is that the fully self-consistent FGWED can capture the Mott-insulating behavior of FeSe.
The strong correlation emerged in the LL by FGWED leads to a Mott transition of FeSe, as shown in Fig.~\ref{Fig_band} (see FGWED\_LL).
The spectral function exhibits a clearly defined energy gap.
It has a direct band gap at $\Gamma$ point.
Both conduction and valence bands are composed of Fe-3$d$ and Se-4$p$ orbitals.
The DOS at $\sim$-2.5 eV is predominantly composed of Se-4$p$ states.
In contrast to the FGWED, the gap was not found within the LGWD framework (see LGWD\_LL).
Instead, a prominent quasi-particle peak appears at the Fermi level.
The peak is attributed mainly to the Fe-3$d$, with a small contribution from the Se-4$p$.

For EL, FeSe is calculated to have metallic bands composed of mainly Fe-3$d$ and small Se-4$p$, by both frameworks, as shown in Fig.~\ref{Fig_band} (see FGWED\_EL and LGWD\_EL).
In LGWD, the spectral function exhibits clear features of hole pockets at $\Gamma$ and electron pockets at M along the M-$\Gamma$ high symmetry line, in agreement with ARPES\cite{laura_prb2016}. 
Whereas, the feature is weakened within the FGWED framework due to stronger correlations, which is evidenced by the Mott-like shape of the DOS in Fig.~\ref{Fig_self} (a).

\begin{figure}[ht]
\centering
\includegraphics[width=0.50
\textwidth]{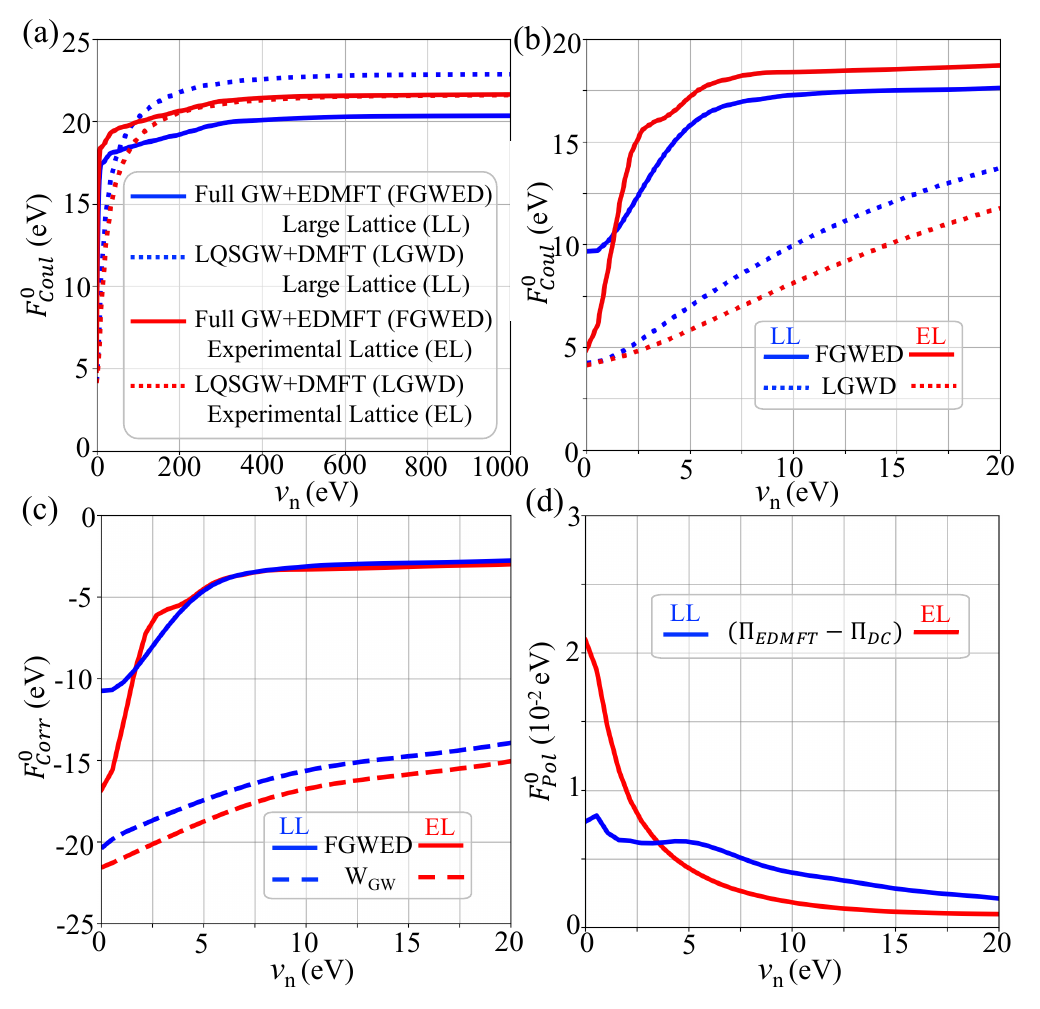}
\caption{\label{Fig_u}
(a) $F^{0}_{Coul}$, the monopole part of the bosonic Weiss field   associated with Fe-3$d$ orbitals within full GW+EDMFT and LQSGW+DMFT for large and experimental lattices. 
(b) Magnified views of $F^{0}_{Coul}$ in the low frequency range. 
(c) $F^{0}_{Corr}$, the correlated monopole part of the bosonic Weiss field associated with Fe-3$d$ orbitals  within FGWED.
(d) $F^{0}_{Pol}$, the monopole part of the embedding polarizability  associated with Fe-3$d$ orbitals  within FGWED.
}
\end{figure}

Figure~\ref{Fig_self} shows DOS and the imaginary part of self-energies of Fe-3$d$ orbitals. 
The self-energies of LL in LGWD and EL in both frameworks display Fermi-liquid-like behavior, resulting in the formation of a quasi-particle peak in the vicinity of the Fermi level.
Table~\ref{table_occ} demonstrates that the occupations of all Fe-3$d$ orbitals in the three cases are away from half-filling, which is a prerequisite for Hund's physics to govern the electronic structure of the quasi-particle in FeSe\cite{byung_lanio2}.
The quasi-particle weight $Z$ of FeSe is found to be lower for the LL than for the EL, within the LGWD framework, as shown in Table~\ref{table_occ}.
This is evidenced by the relatively narrow peak in the DOS, as shown in Fig.~\ref{Fig_self} (a).
For EL, FeSe exhibits similar $Z$ and behavior of self-energy in both frameworks. In contrast, for LL, a strong divergence appears in the self-energies, except for $d_{z^{2}}$, which is fully occupied within FGWED, as shown in Table~\ref{table_occ}. 
Previous works\cite{swagata_sym2021,subhasish_prl2017,byung_lanio2,haul_njp2009,chang_npjcm2020,antonie_arcmp2013,yin_prb2012} have demonstrated that the Hundness is sustained in FeSe. However, increasing the lattice size can suppress the Hundness and lead to Mott physics governing FeSe. Notably, this was only captured within the FGWED framework.

\begin{figure}[ht]
\centering
\includegraphics[width=0.5 \textwidth]{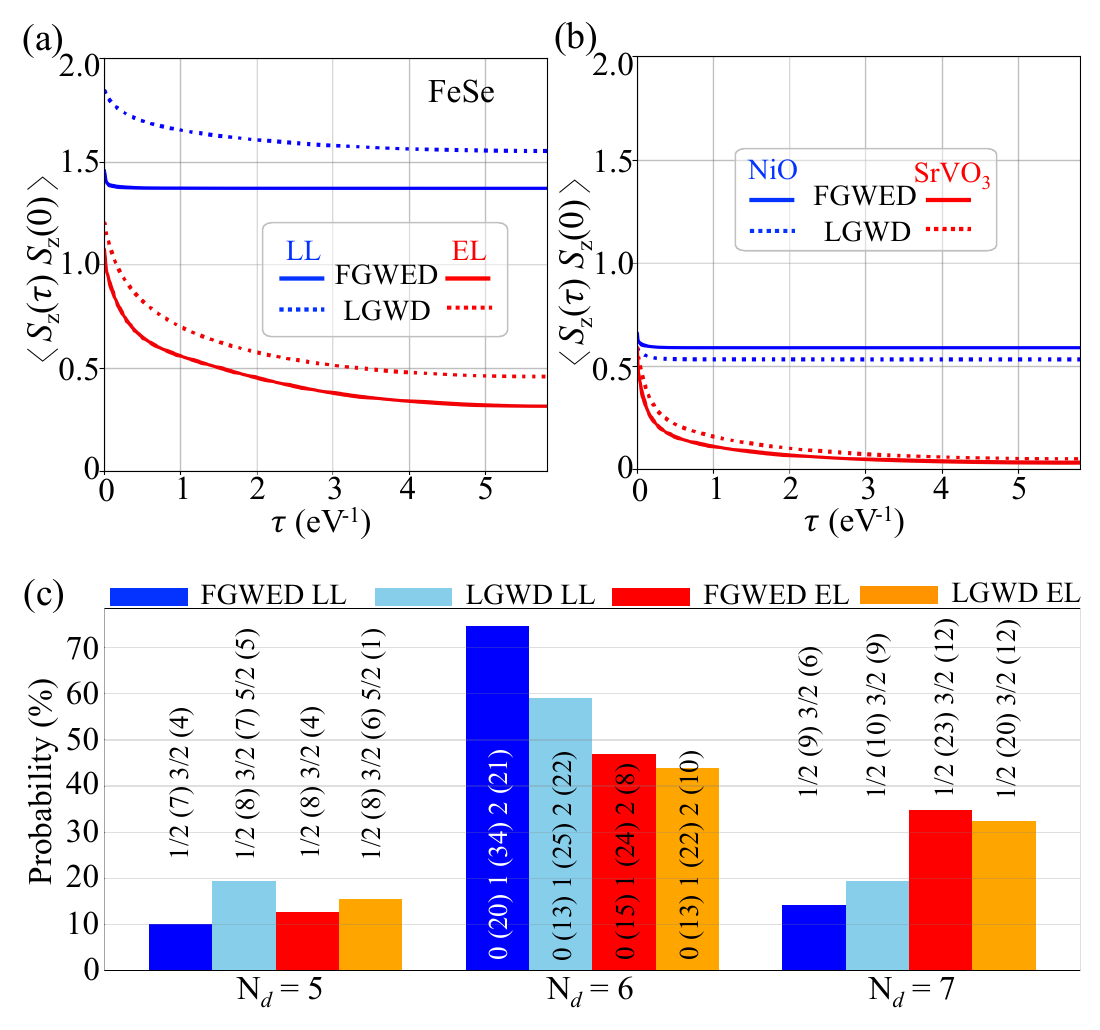}
\caption{\label{Fig_sz}
Local spin moment correlation functions in the imaginary time $\tau$ for Fe-3$d$ within full GW+EDMFT and LQSGW+DMFT for large and experimental lattices.
(b) Local spin moment correlation functions in the imaginary time $\tau$ for Ni-3$d$ in NiO and V-3$d$ in SrVO$_3$ within full GW+EDMFT and LQSGW+DMFT.
(c) Valence histograms for the Fe-3$d$ within full GW+EDMFT and LQSGW+DMFT for large and experimental lattices.
The probability of $S_z$ of spin multiplet configurations are presented in the corresponding valency configuration.
}\end{figure}

\textit{Parameter free many-body methods.}
In order to elucidate the relation between the lattice extension and the Mott gap, which is dependent on the methods employed, we provide a concise workflow of the two frameworks in Figure~\ref{Fig_diagram}.
In FGWED, the $\Psi$ functional is presented as a summation of all two-particle irreducible diagrams generated from the Green's function $G$ and the screened Coulomb interaction $W$.
The Baym-Kadanoff functionals $G$ and $W$\cite{almbladh_VariationalTotalEnergies_1999,chitra_EffectiveactionApproachStrongly_2001} can be employed to approximate the $\Psi$ functional. 
The first order approximation of $W$ is equivalent to the GW approximation of $\Psi^{GW}(G,W)= -\frac{1}{2}\mathrm{Tr}GWG$\cite{almbladh_VariationalTotalEnergies_1999}.
Here, for the higher order calculations, the extended DMFT (EDMFT) approach  was applied to obtain the local approximation of $\Psi(\widetilde{G},\widetilde{W})$\cite{si_KosterlitzThoulessTransitionShort_1996,sengupta_NonFermiliquidBehaviorSpinglass_1995,henrikkajueter_InterpolatingPerturbationScheme_1996} for $d-$ or $f-$ subshell orbitals.
Then, the $\Psi$ functional of GW+EDMFT can be obtained by combining the GW and EDMFT diagrammatic approaches and subtracting the double-counted local $GW$ diagram:
\begin{equation}
  \begin{split}
    \Psi^{GW+EDMFT}(G,W)=  -\frac{1}{2}\mathrm{Tr} GWG+\Psi(\widetilde{G},\widetilde{W})\\
    +\frac{1}{2}\mathrm{Tr} \widetilde{G}\widetilde{W}\widetilde{G},
  \end{split}
\end{equation}
where the quantity $X$ is projected onto the correlated subspace, denoted by $\widetilde{X}$.
Then, the GW+EDMFT $\Psi$ functional, composed of all two-particle irreducible diagrams, is solved self-consistently, as illustrated in Fig.~\ref{Fig_diagram} (a).

Figure~\ref{Fig_diagram} (b) demonstrates that LQSGW+DMFT is a simplified version of full GW+EDMFT and does not reach full self-consistency.
The quasi-particle Hamiltonian is augmented with a one-shot correction from DMFT by combining the DMFT with the LQSGW\cite{Choi_Kotliar-FirstprinciplesTreatment-NpjQuantumMater.-2016,choi2019comdmft}.
The distinguishing factor between the two methods is the bosonic Weiss field $\widetilde{\mathcal{U}}$. While it is calculated through the fully self-consistent GW+EDMFT approach, it is obtained only within constrained random phase approximation (cRPA) \cite{aryasetiawan2004frequency,aryasetiawan_calculations_2006} in the LQSGW+DMFT framework.

\textit{Dynamically screened Coulomb interaction.}
Fig.~\ref{Fig_u} (a) and (b) present the monopole part of the $\widetilde{\mathcal{U}}$ associated with Fe-3$d$ orbitals, $F^{0}_{Coul}$.
The comparison between LL and EL within the LGWD framework reveals that the static $F^{0}_{Coul}(\omega=0)$ values are nearly identical (see Fig.~\ref{Fig_u} (b)). However, the stronger frequency dependent $F^{0}_{Coul}$ seen in Fig.~\ref{Fig_u} (b) results in a reduced band width\cite{casula_prl2012} for LL, which in turn leads to a strong quasi-particle peak in the vicinity of the Fermi level.
For EL, the band width by FLWED is narrower than that by LGWD.
This is attributed to the slightly larger static $F^{0}_{Coul}(\omega=0)$ and more frequency-dependent dynamically screened Coulomb interactions\cite{casula_prl2012}.

At high frequency (Fig.~\ref{Fig_u} (a)), the bare $F^{0}_{Coul}(\omega=\infty)$ by FGWED and LGWD for EL are almost identical. However, for LL, $F^{0}_{Coul}(\omega=\infty)$ by FGWED is smaller than that by LGWD.
The bare Coulomb interaction is enhanced, as the correlated atomic orbitals are localized\cite{liu_2023}.
We tested the Wannier function spreads for all Fe-3$d$ orbitals of LL and EL. These by two methods were comparable, however, they became larger by FGWED only for LL, which reduces $F^{0}_{Coul}(\omega=\infty)$, as evidenced in supplementary Table III.
$F^{0}_{Coul} (\omega)$ in the low frequency region are compared between LL and EL in Fig.~\ref{Fig_u} (b). 
It is noted that FGWED gives the stronger frequency dependence of $F^{0}_{Coul}$ and largest static $F^{0}_{Coul}(\omega=0)$ $\sim$ 10 eV for LL, resulting in diverse self-energy and Mott gap.

Figure~\ref{Fig_u} (c) displays the correlated monopole part of the $\widetilde{\mathcal{U}}$ associated with Fe-3$d$ orbitals, $F^{0}_{Corr}$, which is obtained by subtracting the bare $F^{0}_{Coul}(\omega=\infty)$ from $F^{0}_{Coul}$ in the framework of the FGWED.
In order to assess the non-local screening effect, the $F^{0}_{Corr}$ of $W_{GW}$ are calculated by setting $\Pi_{EDMFT}=\Pi_{DC}=0$  (see Fig.~\ref{Fig_diagram} (a)). 
The $F^{0}_{Corr}$ of $W_{GW}$ exhibits a reduced screening effect for LL compared to EL, resulting in $\sim$ 1 eV larger correlation. 
In addition, the strong correlations observed in the form of large static $F^{0}_{Coul}(\omega=0)$ can also be attributed to the embedding polarizability ($\Pi_{EDMFT}-\Pi_{DC}$), which is associated with local screening effects.
Fig.~\ref{Fig_u} (d) presents the monopole part of the embedding polarizability associated with Fe-3$d$ orbitals, $F^{0}_{Pol}$, within FGWED.
The static $F^{0}_{Pol}(\omega=0)$ for EL is larger than that of LL, resulting from a higher screening effect. 
These results show that the screening effect of both local and non-local interactions on the Fe-3$d$ orbitals is reduced in LL, resulting in the largest static $F^{0}_{Coul}(\omega=0)$.

\textit{Dynamic spin configurations.}
The local spin moment correlation functions $\chi_{S_Z}(\tau)=\langle S_z(\tau)S_z(0)\rangle$ are presented in Fig.~\ref{Fig_sz}. These correlation functions can be used to evaluate the extent of magnetic moment localization\cite{belozerow_prb2023}.
In FGWED, the $\chi_{S_Z}(\tau)$ of FeSe in LL is found to remain constant with an increase of $\tau$, suggesting the presence of strong local magnetic moments.
The localization of electrons due to the Mott transition is analogous to that observed in NiO, a charge transfer Mott insulator\cite{kang_nio}. 
Interestingly, the constant $\chi_{S_Z}(\tau)$ associated with the Mott transition are captured within both frameworks in NiO, as illustrated in Fig.~\ref{Fig_sz} (b). 
For FeSe in the LL within LGWD and in the EL within both frameworks, $\chi_{S_Z}(\tau)$ at $\tau\xrightarrow{} \beta/2$ exhibits a saturated values with different renormalizations, indicating that the Fe-3$d$ electrons are localized (not completely screened) and thus a frozen magnetic moment is present.
This is also one of the indications of Hundness in FeSe\cite{byung_lanio2}.
The behavior of $\chi_{S_Z}(\tau)$ is comparable to that of SrVO$_3$, except that the correlation function saturates to zero in SrVO$_3$, as shown in Fig.~\ref{Fig_sz} (b). This suggests that the magnetic moments are completely screened in SrVO$_3$, unlike FeSe.
The increasing correlation functions from EL to LL of FeSe is reminiscent of the behavior of local spin susceptibility of Na(Fe,Cu)As, where Cu doping leads to a Mott insulator with a $d$-$d$ energy gap of 0.2 eV\cite{skornyakov_prb2021}.
At high Cu-doping levels, $\chi_{S_Z}(\tau)$ remains nearly constant, with large instantaneous local magnetic moments, suggesting the formation of fluctuating local magnetic moments of Fe-3$d$ electrons.

Figure~\ref{Fig_sz} (c) shows the valence histogram for the Fe-3$d$.
The Fe-3$d$ orbitals of FeSe in the EL exhibit a high degree of multi-valency, with Fe-3$d$ occupancy (N$_d$) ranging from 5 to 7 in both frameworks.
The charge transfer between Fe and Se can be attributed to the mixed configurations of $d^6$ and $d^7\underline{L}$, where $\underline{L}$ denotes a ligand hole.
Fig.~\ref{Fig_band} and Fig.~\ref{Fig_self} demonstrate a sizable Se-4$p$ DOS near the Fermi level and a metallic nature of all five Fe-3$d$ orbitals, respectively. This indicates that the multivalency of the system is likely due to charge transfer from the ligand to the Fe-3$d$ orbitals in a configuration interaction similar to that seen in nickel-based compounds\cite{hepting_nature2020,byung_lanio2,kang_nio}.
The smallest yet substantial multivalency of insulating FeSe (LL) within the framework of FGWED indicates that charge transfer in the Mott phase is responsible for the observed valency fluctuations.
The probability of $S_z$ of spin multiplet configurations in the corresponding valency configuration is presented in Fig.~\ref{Fig_sz} (c).
The spin-aligned ($S_z$=5/2) state was observed in N$_d$=5 of FeSe in the LL within the LQSGW framework. This is a significant contribution to the highest spin correlation function of FeSe in the LL, as shown in Fig.~\ref{Fig_sz} (a).
The larger $S_z$ in LL with N$_d$=6 within the both frameworks indicate that an enhancement of Hund's coupling leads to a significant local magnetic moment in FeSe in larger lattices.

\textit{Conclusion.}
The self-consistent embedding of the local impurity self-energy into the non-local GW treatment of the screened Coulomb interaction and self-energy is at the core of fully self-consistent GW+EDMFT~\cite{biermann_prl2003,sun2002extended,kangfgwedmft}.
FeSe has been identified as a Hund metal due to the presence of Fe-3$d$ bands in the vicinity of the Fermi level, where the strong correlations are mainly attributed to $J$.
As the dimensionality of FeSe is reduced to 2D by increasing the interlayer distance,
both local and non-local screening effects on the Fe-3$d$ orbitals decrease.
This introduces a competition between $U$ and $J$ that was previously unidentified.
Only fully self-consistent GW+EDMFT captures the overwhelming $U$ over $J$ in 2D FeSe in its paramagnetic phase, which results in a significant band gap, the emergence of strong local magnetic moments, and a reduced valency fluctuation.

Our findings suggest that FeSe exhibits a Mott insulator transition (MIT) when transitioning from a bulk to a low-dimensional monolayer or nano-structure, providing an explanation for the observed semiconducting behavior.
Consequently, FeSe emerges as a promising quantum material with tunable electron correlations and versatile electronic and magnetic properties, making it a compelling candidate for applications in advanced quantum devices.
Our approach establishes a framework for discovering the electronic structure of FeSe in relation to its atomic geometry without relying on presumptions. 
Further investigations, particularly from the perspective of Mott physics, hold the potential to shed light on the mechanism behind high-temperature superconductivity in monolayer FeSe.

\section*{Acknowledgments} 
This research was supported in part through the use of Information Technologies (IT) resources at the University of Delaware, specifically the high-performance computing resources.
C.H. Park and M. Kim acknowledge the support of the National Research Foundation of Korea (NRF) grant (Grant No. NRF-2022R1A2C1005548). 
The DFT calculation was performed by using high-performance computing clusters in the Quantum Matter Core-Facility (QMCF) of Pusan National University.
A. Janotti was supported by the NSF through the UD-CHARM University of Delaware Materials Research Science and Engineering Center (MRSEC) Grant No. DMR-2011824. This research also used resources of the National Energy Research Scientific Computing Center (NERSC), a U.S. Department of Energy Office of Science User Facility located at Lawrence Berkeley National Laboratory, operated under Contract No. DE-AC02-05CH11231 using NERSC award BES-ERCAP m3340, and the DARWIN computing system: DARWIN – A Resource for Computational and Data-intensive Research at the University of Delaware (NSF MRI award OAC-1919839).

\bigskip
\textbf{Competing Interests} The authors declare no competing interests.

\bigskip
\textbf{Data availability} The data that support the ﬁndings of this study are available from the corresponding
authors upon reasonable request.

\bigskip
\textbf{Author contributions} B.K. and C.P. designed the project. B.K. performed the GW+EDMFT and LQSGW+DMFT calculations and conducted the data analysis. M. K. conducted DFT calculations. All authors wrote the manuscript, discussed the results, and commented on the paper.

\bibliography{ref}

\end{document}